\documentclass[12pt,preprint]{aastex}

\shortauthors{Zhang and Flyer} \shorttitle{Dependence of Helicity
Bound on Boundary Condition}

\begin{document}

\title{The dependence of the helicity bound of force-free magnetic fields on
boundary conditions}

\author{Mei Zhang\altaffilmark{1,2} \& Natasha Flyer\altaffilmark{3}}

\altaffiltext{1}{National Astronomical Observatory, Chinese Academy
of Sciences, 20A Datun Road, Chaoyang District, Beijing 100012,
China} \altaffiltext{2}{High Altitude Observatory, National Center
for Atmospheric Research, PO Box 3000, Boulder, CO 80307, USA}
\altaffiltext{3}{Institute for Mathematics Applied to the
Geosciences, National Center for Atmospheric Research, PO Box 3000,
Boulder, CO 80307, USA}

\begin{abstract}
This paper follows up on a previous study showing that in an open
atmosphere such as the solar corona the total magnetic helicity of a
force-free field must be bounded and the accumulation of magnetic
helicity in excess of its upper bound would initiate a
non-equilibrium situation resulting in an expulsion such as a
coronal mass ejection (CME). In the current paper, we investigate
the dependence of the helicity bound on the boundary condition for
several families of nonlinear force-free fields. Our calculation
shows that the magnitude of the helicity upper bound of force-free
fields is non-trivially dependent on the boundary condition. Fields
with a multipolar boundary condition can have a helicity upper bound
ten times smaller than those with a dipolar boundary condition when
helicity values are normalized by the square of their respective
surface poloidal fluxes. This suggests that a coronal magnetic field
may erupt into a CME when the applicable helicity bound falls below
the already accumulated helicity as the result of a slowly changing
boundary condition. Our calculation also shows that a monotonic
accumulation of magnetic helicity can lead to the formation of a
magnetic flux rope applicable to kink instability. This suggests
that CME initiations by exceeding helicity bound and by kink
instability can both be the consequences of helicity accumulation in
the corona. Our study gives insights into the observed associations
of CMEs with the magnetic features at their solar surface origins.
\end{abstract}



\section{Introduction}

Magnetic helicity is a physical quantity that measures the
topological complexity of a magnetic field, such as the degree of
linkage and/or twistedness in the field (Moffatt 1985, Berger \&
Field 1984). In a previous paper (Zhang et al. 2006, hereafter
referred to as ZFL) we proposed that in an open atmosphere such as
the solar corona there is an upper bound on the total magnetic
helicity that a force-free field can contain. The accumulation of
magnetic helicity in excess of this upper bound would initiate a
non-equilibrium situation, resulting in a coronal mass ejection
(CME) as a natural product of coronal evolution.

Our approach (Zhang \& Low 2003, 2005, ZFL) shifts the traditional
focus on the mechanism for CME evolution from the storage of
magnetic energy to the accumulation of magnetic helicity, although
the two types of considerations are not necessarily exclusive with
each other. The advantage of using magnetic helicity as a more
fundamental physical quantity is that from observations we know that
the magnetic fields are emerging from the solar interior with a
preferred helicity sign in each solar hemisphere (Pevtsov et al.
1995, Rust \& Kumar 1996, Bao \& Zhang 1998, Zhang 2006). As a
result, an accumulation of the total magnetic helicity in the corona
becomes unavoidable because the total magnetic helicity is
approximately conserved in the corona during coronal processes
including fast magnetic reconnection (Berger 1984).

In this paper, we study how the magnitude of the upper bound of the
total magnetic helicity depends on the boundary condition. Section 2
presents the model with two new boundary conditions. Results and
analysis are presented in Section 3. In Section 4, a brief summary
of the paper is given.


\section{The Model}

\subsection{The governing equation}

Following Flyer et al. (2004) and ZFL, we use the families of
power-law axisymmetric force-free fields to understand the basic
physical properties of interest.

With axisymmetry, the solenoidal magnetic field ${\bf B}$ in
$r > 1$ can be written in the form of
\begin{equation}
{\bf B} = {1 \over r \sin \theta} \left[
{1 \over r}{\partial A \over \partial \theta} ~,
~ - {\partial A \over \partial r}~, ~Q (A) \right] ~,
\end{equation}
\noindent where the flux function $A$ defines the poloidal magnetic
field and the function $Q$ defines the toroidal (or azimuthal)
field.

$Q$ is defined as a strict power-law in $A$ with the form
\begin{eqnarray}
Q^2 (A) = \frac{2 \gamma}{n+1} A^{n+1} ~,
\end{eqnarray}
\noindent where $n$ is an odd constant index required to be no less
than 5 in order for the field to possess finite magnetic energy in
$r > 1$ and $\gamma$ is a free parameter which we choose to be
positive without loss of generality. This form of $Q$ reduces the
force-free condition to the following governing equation for the flux
function $A$:
\begin{eqnarray}
{\partial^2 A \over \partial r^2} + {1 - \mu^2 \over r^2}
{\partial^2 A \over \partial \mu^2} + \gamma A^n = 0 ~~.\label{eqnA}
\end{eqnarray}

This governing equation was solved numerically as a boundary value
problem within domain $r > 1$ in Flyer et al. (2004) and ZFL,
subject to the prescribed boundary flux distribution of
\begin{equation}
A|_{r = 1} = \sin^2 \theta ~~.
\label{dipolar}
\end{equation}
\noindent This boundary flux distribution and its associated normal
field distribution are plotted in the top panels of Figure 1. All
force-free fields discussed in Flyer et al. (2004) and ZFL share the
boundary flux distribution given by (\ref{dipolar}), which is the
same as that for a dipole potential field corresponding to the
solution of equation (\ref{eqnA}) with $\gamma=0$. The solutions of
(\ref{eqnA}) with boundary condition (\ref{dipolar}) shall be
referred to as dipolar force-free fields or dipolar fields for
short. We refer interested readers to Flyer et al. (2004) and ZFL
for various properties of dipolar force-free fields for the cases of
$n = 5, 7, 9$.


\subsection{Two new boundary conditions}

In this paper, we solve the same governing equation (\ref{eqnA})
using the numerical methods described in Flyer et al. 2004, 2005
but subject to two new and distinctively different boundary
conditions. We use these numerical solutions to investigate how
properties we discussed in ZFL would change with the new boundary
conditions.

The first new boundary condition has the flux distribution of
\begin{equation}
A|_{r = 1} = \sin^{12}\theta ~~.
\end{equation}
\noindent This flux distribution and its associated normal field
distribution are plotted in the middle panels of Figure 1. We see
that this new boundary condition has its flux concentrated nearer to
the equator than that of the dipolar field. This makes it more like
solar active regions with its normal field strength much higher at
equatorial regions than that near the poles. We shall refer this
family of power-law axisymmetric force-free fields as bipolar
force-free fields or bipolar fields hereafter.

The second new family of solutions are those that we shall refer to
as multipolar force-free fields or multipolar fields. They are also
the solutions to (\ref{eqnA}) but subject to the following boundary
condition:
\begin{equation}
A|_{r = 1} = \sin^2\theta \times (1- 5 \times \cos^2\theta) ~~~.
\end{equation}
\noindent Its flux distribution and associated normal field
distribution are plotted in the bottom panels of Figure 1. We see
that fields with this boundary condition have both positive and
negative magnetic fields in one hemisphere, distinctively different
from those in dipolar force-free fields and bipolar force-free
fields where fields in one hemisphere have the same magnetic
polarity.

Table 1 presents a comparison of the three boundary conditions. We
see that their common feature is that they all have $A|_{r=1}=0$ at
solar poles and $A|_{r=1}=1$ at the equator. The dipolar boundary
and bipolar boundary differ by the contrast of their respective
maximum $B_r|_{r=1}$ values to their $B_r|_{r=1}$ values at the
northern pole. The multipolar boundary condition differs from the
dipolar and bipolar boundary conditions by the existence of both
magnetic polarities in each hemisphere, indicated by its negative
minimum $A$ value.

In Table 1 we also present the total surface poloidal flux ($F_p$)
for the three boundary conditions, where
\begin{equation}
F_p = \int_{r=1} B_r(>0) ds = 2\pi\int^{\frac{\pi}{2}}_0 |B_r|
\sin\theta d\theta ~~~.
\end{equation}
\noindent We see that dipolar and bipolar fields have the same total
surface poloidal flux ($2\pi$) and the multipolar fields have a
larger total surface poloidal flux ($5.2\pi$). In the latter
development of the paper, we will normalize calculated magnetic
helicity by $F_p^2$ in order to make fields with different boundary
conditions comparable.


\section{Results and Analysis}

As in ZFL, for each new boundary condition, we numerically solve
(\ref{eqnA}) for three cases: $n=5$, $n=7$ and $n=9$. In each case, the
numerical method, that is, the Newton's iteration combined with a
pseudo-arc length continuation scheme, guarantees the completeness
of each solution branch generated by the $\gamma$ values.

Also as in ZFL, for each obtained solution, we calculate three
physical quantities of the field. They are: total magnetic energy
\begin{equation}
E  =   \int_{r > 1} {B^2 \over 8 \pi} dV
     = \frac{1}{4} \int_{r = 1}
    \left( B_r^2  - B_{\theta}^2 - B_{\varphi}^2
        \right) \sin \theta d\theta ~~,
\end{equation}
\noindent total azimuthal flux
\begin{equation}
F_{\varphi} = \int_{r > 1} |B_{\varphi}| r dr d\theta =
\sqrt{\frac{\gamma}{m+1}} \int_{r > 1} |A|^{m+1} {dr d\theta \over
\sin \theta} ~~,
\label{Fphi}
\end{equation}
\noindent and total relative magnetic helicity
\begin{equation}
H_R = 4 \pi \int_{r>1}  A B_{\varphi} r dr d \theta = 4 \pi
\sqrt{\frac{\gamma}{m+1}} \int_{r > 1} A^{m+2} {dr d\theta \over
\sin \theta} ~.
\end{equation}
\noindent The derivation of these formula can be found in ZFL. The
only difference is that since we are also considering multipolar
fields the absolute value of $B_r$ (that is, $|B_r|$) is introduced
when calculating total azimuthal flux.

In the geometric simplicity of these force-free fields, the
equilibrium in each case is due to the magnetic tension force of the
poloidal flux confining the magnetic pressure of the azimuthal flux.
The magnitude of the tension force is sensitively related to the
poloidal flux at the inner boundary ($r=1$) that serves as an anchor
agent. Moreover, the poloidal flux and its tension force become
weaker in the outward radial direction. Hence, if the azimuthal flux
becomes too large it can not be confined by such a mechanism as was
shown in ZFL for dipolar fields. The magnetic pressure is
independent of the sign of the field. Therefore, the total unsigned
azimuthal flux given by (\ref{Fphi}) is more relevant for the
consideration of flux confinement than the total signed azimuthal
flux.


\subsection{Helicity upper bound of bipolar fields}

Figure 2 presents the variations of the total magnetic energy ($E$),
total azimuthal flux ($F_{\varphi}$) and total magnetic helicity
($H_R$) versus $\gamma$ along the solution curve for bipolar force-free
fields with $n=5$, $n=7$ and $n=9$. The figure is similar to the
Figure 2 of ZFL except that Figure 2 in ZFL is for dipolar
force-free fields. Each point along the solution curve, denoted by a
plus symbol in the figure, represents a solution to (\ref{eqnA}). By
solution curve we mean that all solutions along the curve are
obtained with the same boundary condition and the same constant
index $n$ but with a monotonically increasing magnitude of total
azimuthal flux.

From these curves of solutions we see that there may be upper bounds
on the total magnetic energy, total azimuthal flux as well as total
magnetic helicity for bipolar fields, as that we have suggested in
ZFL for dipolar fields.

From Figure 2 we can also see that the magnitude of the helicity
upper bound for bipolar fields is smaller than that for dipolar
fields. While the upper bound for dipolar fields is close to 14, the
upper bound for bipolar fields only approaches 9. To illustrate this
further, we plot in Figure 3 the variation of the total magnetic
helicity ($H_R$) versus azimuthal flux ($F_{\varphi}$) along the
solution curve for fields with the dipolar (top panels) and bipolar
(bottom panels) boundary conditions. Here we have normalized the
values of total magnetic helicity ($H_R$) of each field by the
square of their corresponding surface poloidal fluxes ($F_p^2$) as
those in Demoulin et al. (2002), van Driel-Gesztelyi et al. (2003)
and Demoulin (2007). We see that while the upper bound of
$H_R/F_p^2$ is about 0.35 for dipolar fields, for bipolar fields it
is significantly lower at 0.22.

This result suggests a dependence of the helicity upper bound on the
boundary condition, which in our view gives an insight into observed
associations between CMEs and magnetic features at their solar
surface origins. Observationally we know that CMEs can be triggered
by flux emergence (e.g. Feynman \& Martin 1995, Subramanian \& Dere
2001) and converging motions (e.g. Martin 1990). Different
theoretical models have also proposed that CME-type eruptions can be
triggered by various surface field variations (e.g. Chen \& Shibata
2000, Amari et al. 2000, 2003a, 2003b).

We suggest that the common physics underlying the different
mechanisms associated with such surface field variations is the
dependence of helicity upper bound on the boundary condition. When a
magnetic field has accumulated a certain amount of magnetic helicity
(but not yet enough for an eruption) then a change of the boundary
condition could lower the helicity upper bound, resulting in a
non-equilibrium situation and hence a CME eruption under the new
boundary condition. For example, if $H_R/F_p^2$ were 0.3 for a
dipolar boundary condition, then an evolutionary change to a bipolar
boundary condition would result in a CME eruption because the
applicable upper bound on the conserved total helicity has been
reduced as suggested by our numerical experiments.

A note to address here is that although in this paper we have
emphasized the role of boundary condition variations this does not
mean that the role of magnetic helicity accumulation becomes less
important. A change of the boundary condition may bring in an
eruption only when the field has accumulated enough helicity for an
eruption under the new boundary condition. If not, the field does
not erupt even when the boundary flux distribution is changing. This
is consistent with the observation (Zhang et al. 2007) that although
flux emergences are indeed found to be associated with CME
eruptions, the same rate of flux emergence can also be found when
there is no CME or solar activities. This means that flux emergence
may be a trigger of a CME eruption, but flux emergence alone do not
guarantee an eruption.

Another interesting result from our calculations is that these normalized
$H_R/F_p^2$ helicities, estimated from simple axisymmetric power-law
force-free fields, lie close to those estimated from observations.
Van Driel-Gesztelyi et al. (2003) and Demoulin (2007) pointed out
that the $H_R/F_p^2$ numbers, estimated from the extrapolated
magnetic fields based on observed photospheric magnetograms, are
between 0.02 to 0.2. Our helicity upper bound numbers of dipolar and
bipolar fields are just a little higher. Notice that the numbers
estimated from the observations may be somehow underestimated
because of the limited spatial resolution of the observed
magnetograms. So there seems to be consistency between the
theoretical $H_R/F_p^2$ helicities and those estimated from
observations.

Figure 4 presents four field configurations selected from the $n=9$
solution curve, positions of which along the curve are illustrated
in Figure 2. We see that starting from the potential field (Panel A)
the curve first reaches the maximum-$\gamma$ field (Panel B) and
then the field with maximum total magnetic energy (Panel C) and
then the one with maximum total azimuthal flux (Panel D). A clear
bubble (representing a flux rope) is presented in the field of Panel
C but not in the field of Panel D where the latter actually
possesses more total azimuthal flux than the former. This tells us
that although the existence of a flux rope (or flux ropes) in the
low corona does represent a storage of a certain amount of magnetic
helicity (see more discussions in Zhang \& Low 2003) it is not
necessary that they are present in the field with a maximum helicity
storage.


\subsection{Helicity upper bound of multipolar fields}

As in Figure 2, we present in Figure 5 the variations of the total
magnetic energy, total azimuthal flux and total magnetic helicity versus
$\gamma$ along the solution curve with $n=5$, $n=7$ and $n=9$ but
for the multipolar fields. We see that these curves also suggest the
existence of upper bounds on the total magnetic energy, total
azimuthal flux as well as total magnetic helicity as those for
dipolar fields and bipolar fields.

The figure also shows that the magnitude of the helicity upper bound
for multipolar fields is smaller than that for dipolar fields. This
reduction is even more evident in Figure 6 where we plot $H_R/F_p^2$
versus $F_\varphi$. We see that the upper bound of $H_R/F_p^2$
for multipolar fields is below 0.04, almost ten times smaller than
that of dipolar fields.

Such a severe reduction of $H_R/F_p^2$ upper bound not only further
confirms our previous result that the helicity upper bound is
dependent on the boundary condition, but also brings our theoretical
$H_R/F_p^2$ value even closer to those estimated from
observations (Regnier et al. 2005). Furthermore, the severe
reduction of the helicity upper bound in terms of $H_R/F_p^2$ values
may also explain why solar eruptions such as CMEs are more likely to
happen in complicated active regions where the multipolar field, by
the above property, will take less time to reach its helicity bound,
producing an eruption.

As in Figure 4, Figure 7 presents four field configurations selected
from the $n=7$ solution curve of the multipolar fields. We see again
a clear bubble in the field of Panel C but not in Panel D. Therefore,
as with bipolar fields, multipolar fields with maximum helicity storage
need not contain a flux rope.


\subsection{Kink instability}

A rope of highly helical field is susceptible to an instability that
causes the rope to kink (Friedberg 1987). From elementary
calculations, this kink instability sets in if a critical twist is
exceeded ($T>T_c$). The exact value of $T_c$ depends on the detailed
field models, and could increase from the traditional
Kruskal-Shafranov limit $T_c=2\pi$ to $T_c=2.5\pi$ (Hood \& Priest
1981) and $T_c=4.8\pi$ (Mikic et al. 1990).

Since we have helical flux tubes (or flux ropes) present in our
solutions, it is interesting to investigate whether these flux ropes
have exceeded the kink instability. Figure 8 presents the variation
of the average twist ($T$) versus $\theta_0$ for two fields. One of the
fields is the bipolar $n=9$ maximum-energy field, the one presented
in Panel C of Figure 4. The other is the multipolar $n=7$
maximum-energy field, the one presented in Panel C of Figure 7. Here
$\theta_0$ is the angle from the equator. The average twist ($T$) is
obtained from $H_R^\prime/(F^\prime_p)^2$, where $H_R^\prime$ is the
total relative magnetic helicity in domain $\Omega^\prime$, enclosed
by the $r=1$ surface and the flux surface with $A =
A|_{r=1,\theta=\theta_0}$, and $F_p^\prime$ is the total surface
($r=1$) polodial flux of the domain $\Omega^\prime$.

We see that in both fields the average twist of the central part of
the field (that is, where the flux rope is located) has exceeded the
kink instability criteria, $T_c=2.5\pi$ of Hood \& Priest (1981) or
$T_c=4.8\pi$ of Mikic et al. (1990). This tells us that with the
accumulation of a certain amount of magnetic helicity, the flux rope
formed in the field can possess a twist number that is larger than
the kink instability criteria. If other necessary conditions are
favorable, for example, if the field has accumulated enough free
magnetic energy, an eruption may happen even before the field has
reached its helicity upper bound state. In that sense, reaching the
helicity upper bound state may not be a necessary condition for
eruption, but the helicity upper bound is a sufficient condition
upon which an eruption will become unavoidable. This also shows that
CME eruptions that are initiated by the kink instability (e.g. Torok
and Kliem 2005, Fan \& Gibson 2007) or by the existence of helicity
upper bound could both be viewed as the consequences of magnetic
helicity accumulation and they are not mutually exclusive.


\section{Conclusion}

In this paper, we continue our study on the hydromagnetic origin of
CMEs in terms of magnetic helicity accumulation. As in a previous
paper (ZFL), we numerically solve (\ref{eqnA}) to get families of
axisymmetric power-law force-free fields, but subject to two new
boundary conditions.

By analyzing and comparing obtained solutions for three different
boundary conditions we conclude the following:

\begin{enumerate}
\item The suggestion that there may be an upper bound on the total magnetic
helicity for force-free fields is also found for the two new
boundary conditions.

\item The magnitude of the helicity upper bound of force-free fields
is non-trivially dependent on the boundary condition. In our
examples, the fields with a surface flux distribution more like a
simple active region (bipolar fields) have their helicity upper
bound smaller than that of fields with dipolar boundary condition.
For multipolar fields, the helicity upper bound ($H_R/F_p^2$) can be
ten times smaller than that of dipolar fields. These results provide
some insights into the observed association of CMEs with flux
emergence and surface field variation. These results also suggest a
physical reason why eruptions are more likely to happen in
complicated active regions.

\item CME initiations by kink instability and by the existence of a
helicity upper bound can both be the result of magnetic helicity
accumulation in the corona. They do not exclude each other.

\end{enumerate}

\acknowledgements

We thank Pascal Demoulin, BC Low, Alexander Nindos and the anonymous
referee for helpful comments. This work was partly supported by the
One-Hundred-Talent Program of Chinese Academy of Sciences, Chinese
National Key Basic Research Science Foundation (G2006CB806300),
Chinese National Science Foundation Grant 40636031, and USA NSF
ATM-0548060 Programs. The National Center for Atmospheric Research
is sponsored by the National Science Foundation. Natasha Flyer would
like to acknowledge the support of NSF grant ATM-0620100.


\clearpage
\begin{table}
\caption{Comparison of the three boundary conditions}

\begin{tabular}{|c|c|c|c|c|c|c|c|c|c|}
\hline
 & \multicolumn{4}{|c|}{$A$ (at $r=1$)} & \multicolumn{4}{|c|}{$B_r$ (at $r=1$)} & $F_p$ \\
\hline
 & north pole & equator & max. & min. & north pole & equator & max. & min. &  \\
\hline
dipolar & 0 & 1 & 1 & 0 & 2 & 0 & 2 & -2 & $2\pi$ \\
\hline
bipolar & 0 & 1 & 1 & 0 & 0 & 0 & 2.25 & -2.25 & $2\pi$ \\
\hline
multipolar & 0 & 1 & 1 & -0.8 & -8 & 0 & 8 & -8 & $5.2\pi$ \\
\hline
\end{tabular}
\end{table}

\clearpage

\begin{figure}
\centerline{\includegraphics[width=160mm]{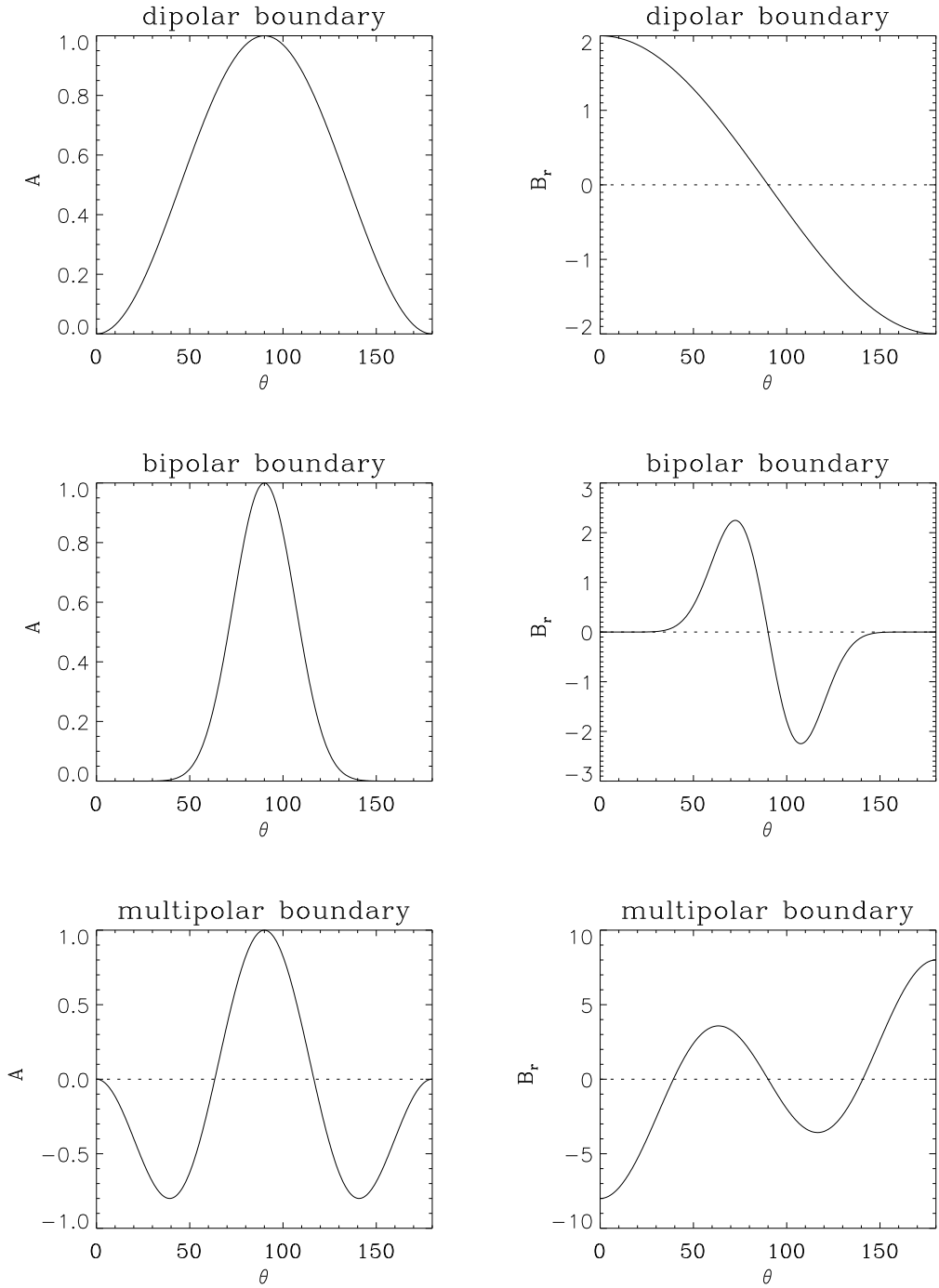}} \caption{Flux
distribution (left panels) and normal field distribution (right
panels) of dipolar fields (top panels), bipolar fields (middle
panels) and multipolar fields (bottom panels) at $r=1$.}
\end{figure}

\begin{figure}
\centerline{\includegraphics[width=130mm]{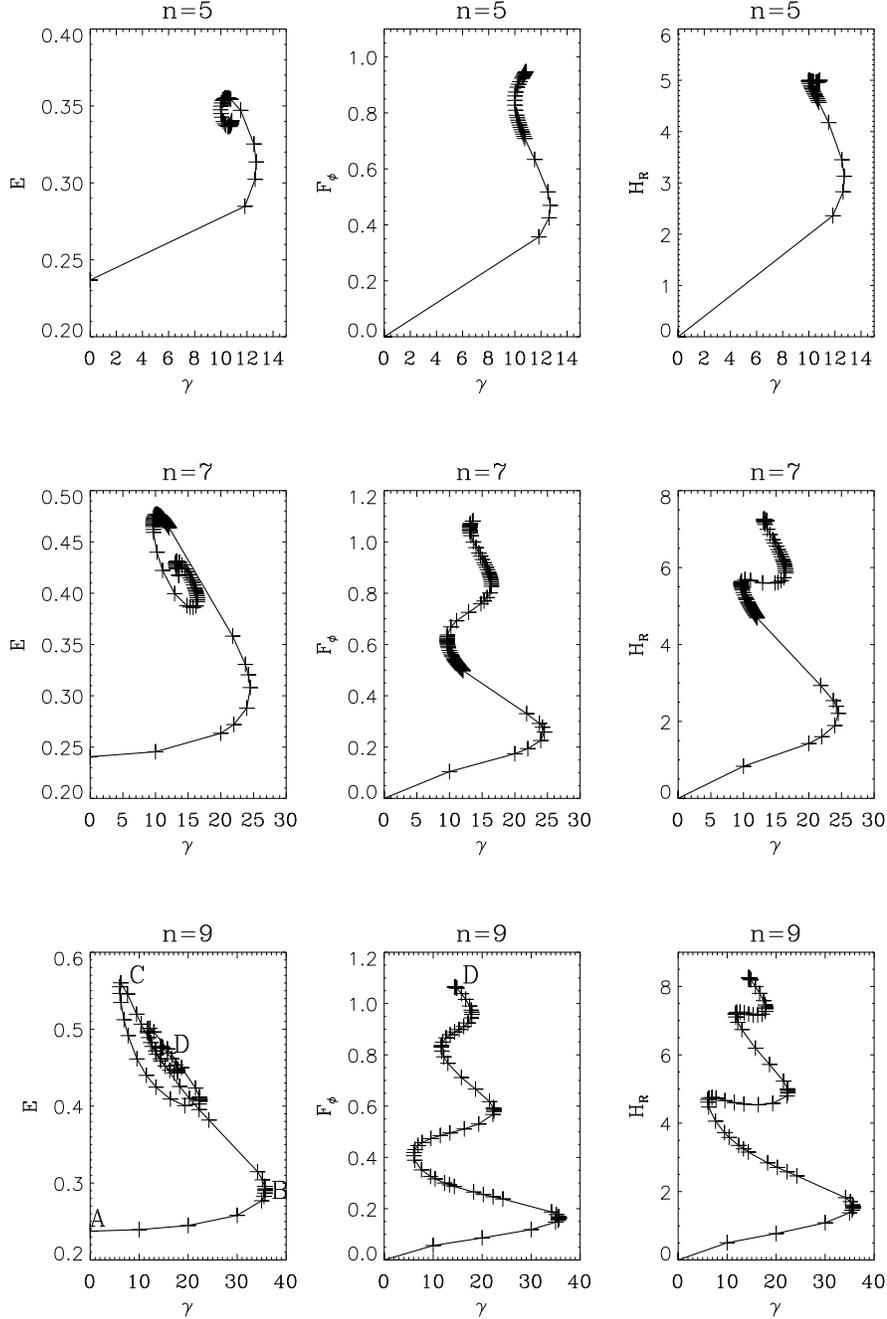}} \caption{Variation of
the total magnetic energy ($E$), azimuthal flux ($F_{\varphi}$) and
total magnetic helicity ($H_R$) vs $\gamma$ along the solution curve
for $n=5$ (top panels), $n=7$ (middle panels) and $n=9$ (bottom
panels) fields with the bipolar boundary condition. Each point,
denoted by a plus symbol in the plot, represents a solution to
(\ref{eqnA}). Letters (A, B, C and D) in the bottom panels indicate
the positions along the solution curve of the four fields plotted in
Figure 4.}
\end{figure}

\begin{figure}
\centerline{\includegraphics[width=160mm]{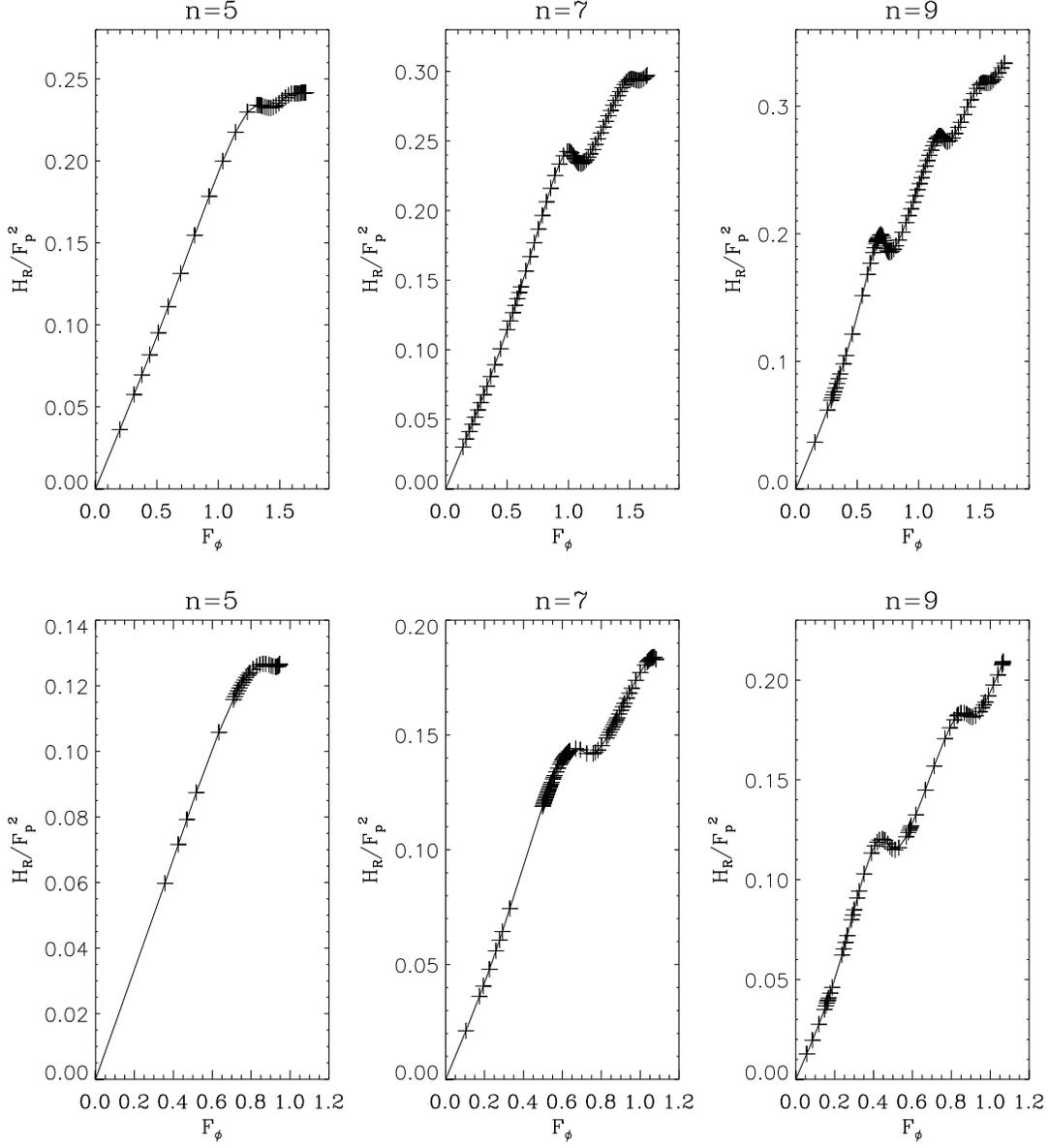}}
\caption{Variation of the total magnetic helicity ($H_R$) vs
azimuthal flux ($F_{\varphi}$) along the solution curve for fields
with the dipolar (top panels) and bipolar (bottom panels) boundary
conditions. Here the total magnetic helicity ($H_R$) of each field
has been normalized by the square of their corresponding poloidal
flux ($F_p^2$).}
\end{figure}

\begin{figure}
\centerline{\includegraphics[width=180mm]{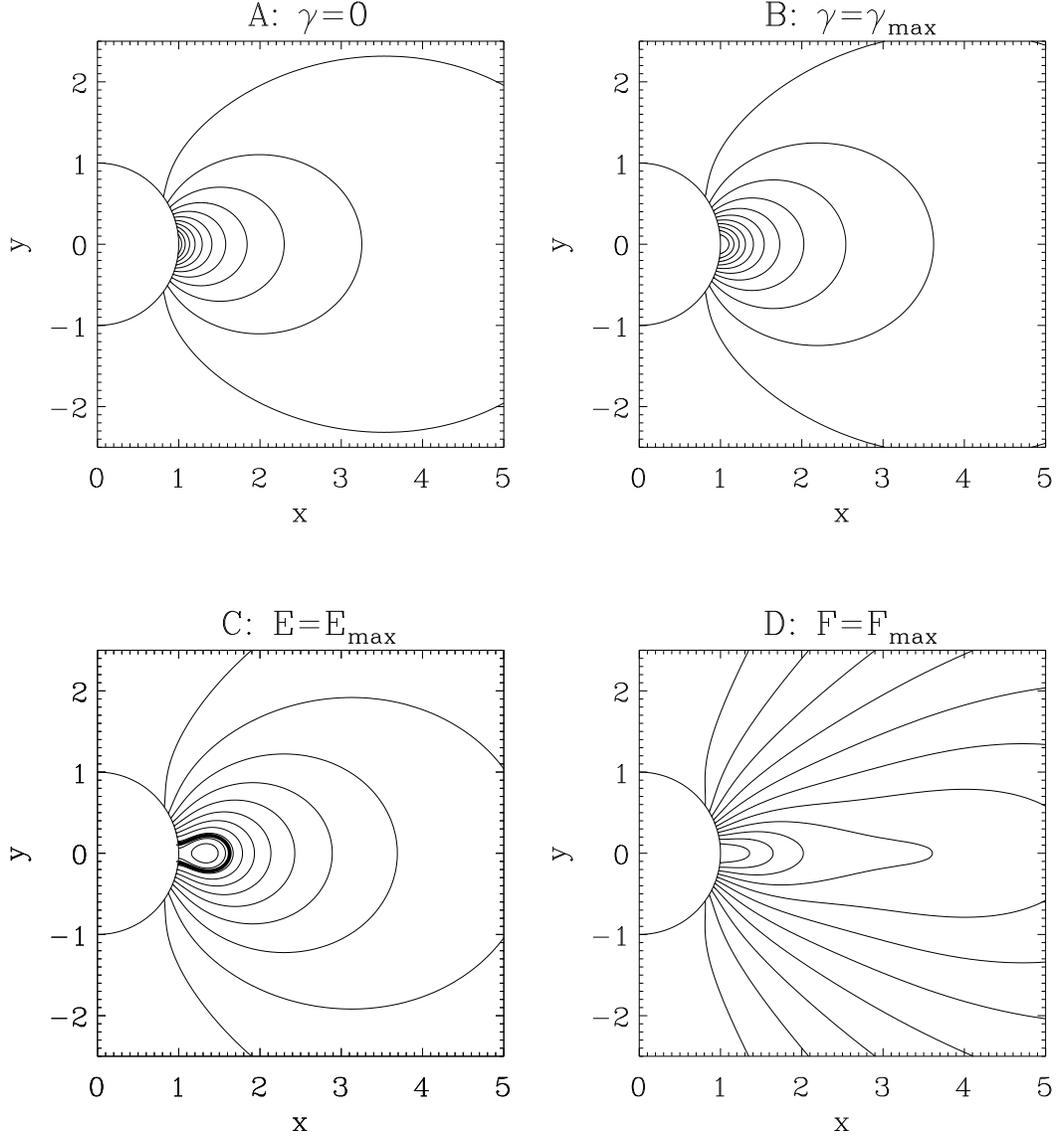}} \caption{Field
configurations of four fields selected from the $n=9$ bipolar
fields. These contours of flux function $A$ represent the lines of
force of the axisymmetric field projected on the $r-\theta$ plane.
Panel A: the potential field; Panel B: the field with the maximum
$\gamma$ value among all $n=9$ solutions; Panel C: the field
possessing the maximum total magnetic energy among all $n=9$
solutions. The thick line in this plot outlines the range within
which the average twist is greater $4\pi$ (see text in Section 3.3
for details); Panel D: the field possessing the maximum azimuthal
flux among all $n=9$ solutions.}
\end{figure}

\begin{figure}
\centerline{\includegraphics[width=150mm]{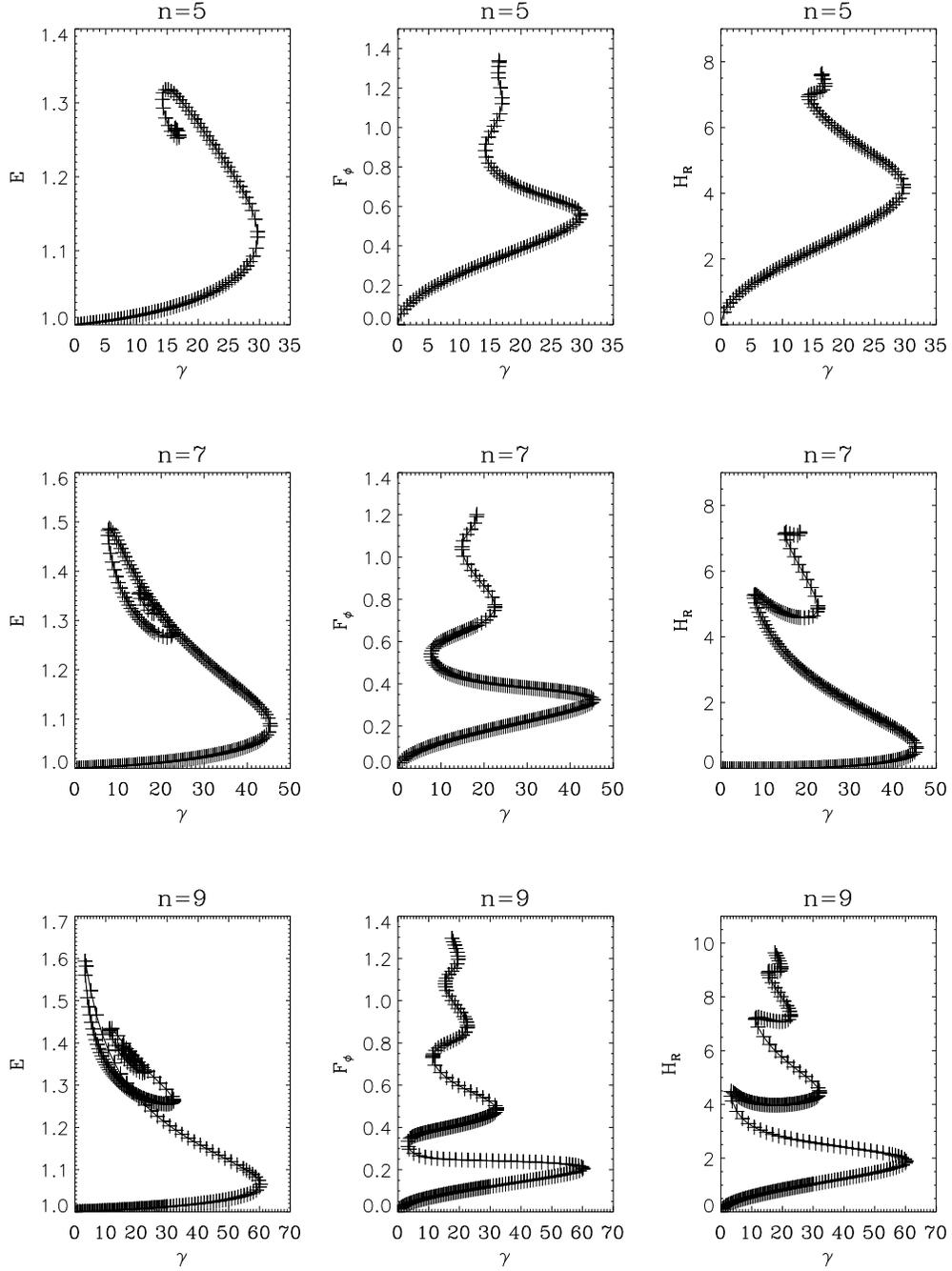}} \caption{Same as
Figure 2, but for fields with the multipolar boundary condition.}
\end{figure}

\begin{figure}
\centerline{\includegraphics[width=160mm]{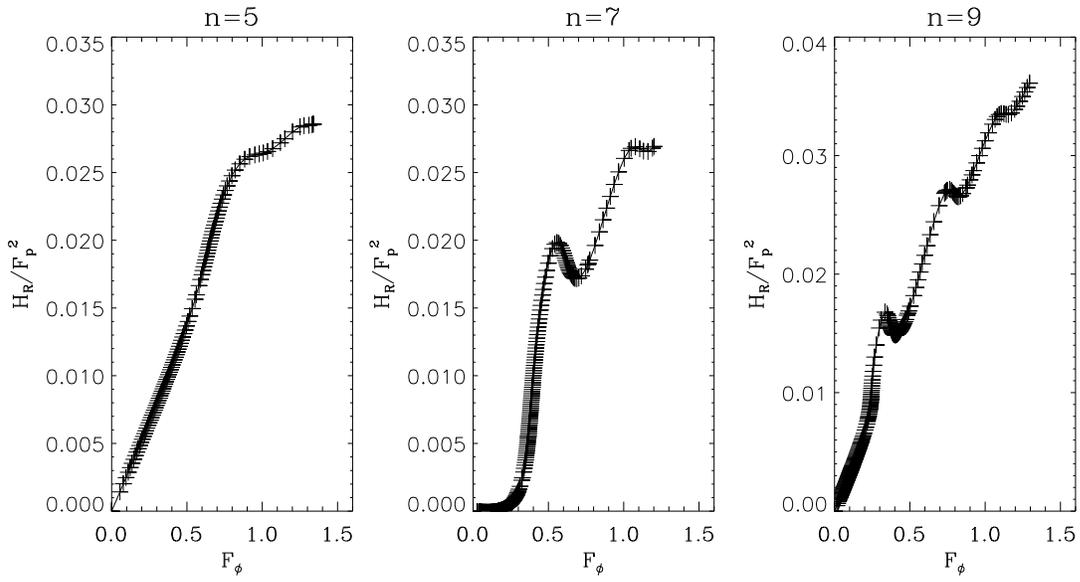}} \caption{Variation of
the total magnetic helicity ($H_R$), normalized by the square of
poloidal flux ($F_p^2$), vs azimuthal flux ($F_{\varphi}$) for
fields with the multipolar boundary condition.}
\end{figure}

\begin{figure}
\centerline{\includegraphics[width=180mm]{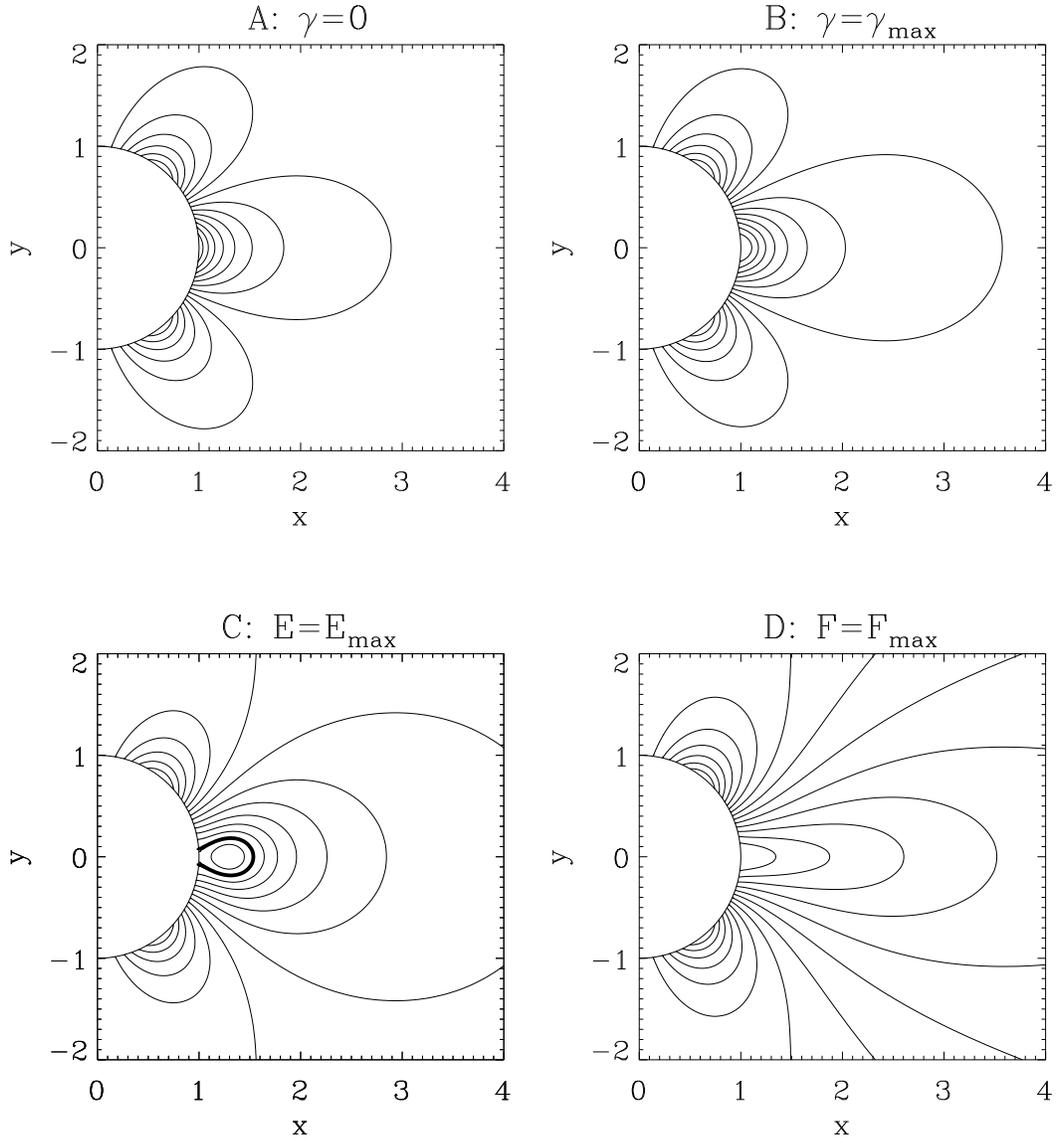}} \caption{Same as
Figure 4, but for the $n=7$ multipolar case.}
\end{figure}

\begin{figure}
\centerline{\includegraphics[width=120mm]{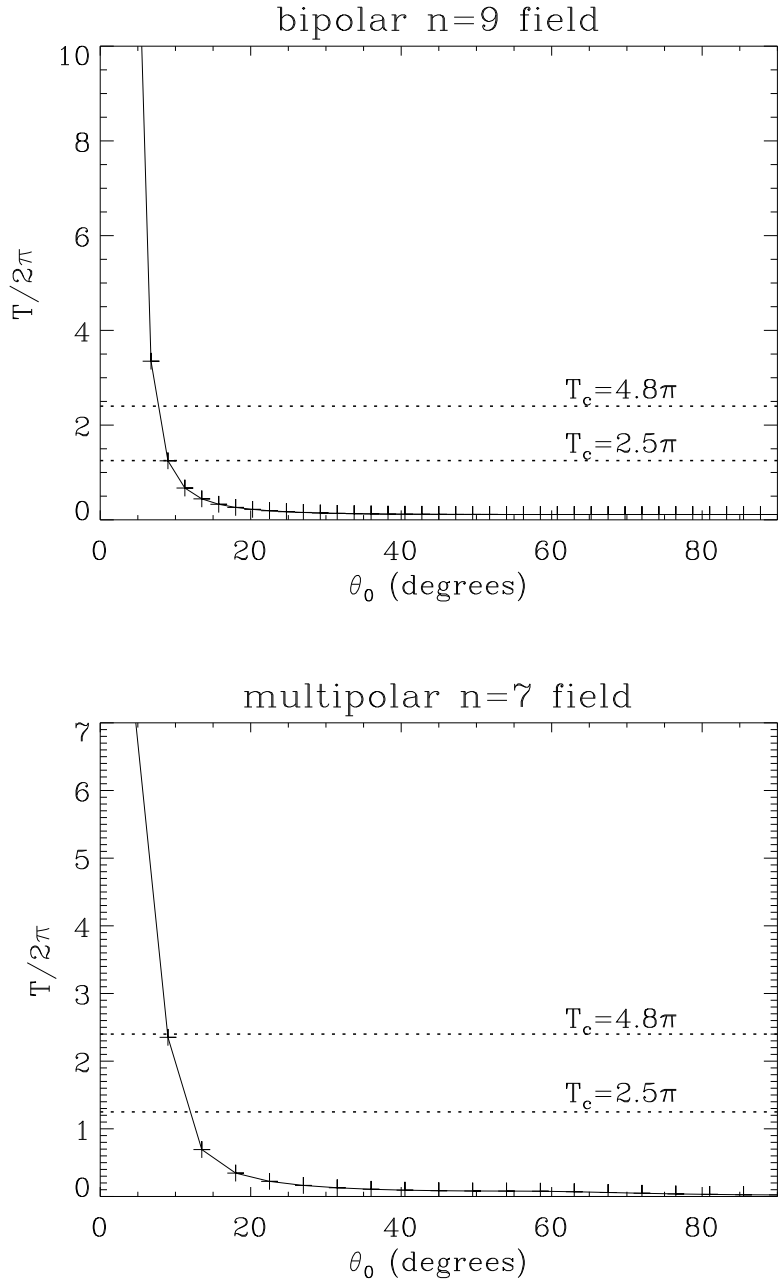}}
\caption{Variation of average twist ($T$) vs $\theta_0$ (angle from
the equator) for the bipolar $n=9$ maximum-energy field (top panel)
and multipolar $n=7$ maximum-energy field (bottom panel). See text
for the calculation of $T$. The dashed lines indicate the critical
twist of kink instability where $T_c=2.5\pi$ (Hood \& Priest 1981)
or $T_c=4.8\pi$ (Mikic et al. 1990). See also Panel C in Figures 4
and 7 for field configurations.}
\end{figure}


\end{document}